\documentclass[10pt]{article}
\usepackage{graphicx}
\usepackage{latexsym}
\usepackage{amssymb}
\usepackage{amsfonts}
\usepackage{amsmath}
\usepackage{theorem}
\newtheorem{theorem}{Theorem}
\newtheorem{lemma}{Lemma}

\newtheorem{definition}{Definition}

\makeatletter
\newcommand{\contraction}[5][1ex]{%
  \mathchoice
    {\contraction@\displaystyle{#2}{#3}{#4}{#5}{#1}}%
    {\contraction@\textstyle{#2}{#3}{#4}{#5}{#1}}%
    {\contraction@\scriptstyle{#2}{#3}{#4}{#5}{#1}}%
    {\contraction@\scriptscriptstyle{#2}{#3}{#4}{#5}{#1}}}%
\newcommand{\contraction@}[6]{%
  \setbox0=\hbox{$#1#2$}%
  \setbox2=\hbox{$#1#3$}%
  \setbox4=\hbox{$#1#4$}%
  \setbox6=\hbox{$#1#5$}%
  \dimen0=\wd2%
  \advance\dimen0 by \wd6%
  \divide\dimen0 by 2%
  \advance\dimen0 by \wd4%
  \vbox{%
    \hbox to 0pt{%
      \kern \wd0%
      \kern 0.5\wd2%
      \contraction@@{\dimen0}{#6}%
      \hss}%
    \vskip 0.5ex
    \vskip\ht2}}

\newcommand{\contraction@@}[3][0.05em]{%
  \hbox{%
    \vrule width #1 height 0pt depth #3%
    \vrule width #2 height 0pt depth #1%
    \vrule width #1 height 0pt depth #3%
    \relax}}
\makeatother

\begin{document}

\title{\bf Overlap Fluctuations from Random Overlap Structures}
\author{Adriano Barra\footnote{King's College London,
Department of Mathematics, Strand, London WC2R 2LS, United Kingdom, and
Dipartimento di Fisica, Universit\`a di Roma
``La Sapienza'' Piazzale Aldo Moro 2, 00185 Roma, Italy,
{\tt<Adriano.Barra@roma1.infn.it>}},
Luca De Sanctis
\footnote{ICTP, Strada Costiera 11, 34014 Trieste, Italy,
{\tt<lde\_sanc@ictp.it>}}}

\def\be{\begin{equation}}
\def\ee{\end{equation}}
\def\bc{\begin{center}}
\def\ec{\end{center}}

\maketitle

\begin{abstract}

We investigate overlap fluctuations of the Sherrington-Kirkpatrick
mean field spin glass model in the framework of the Random Overlap
Structure (ROSt). The concept of ROSt has been introduced recently
by Aizenman and coworkers, who developed a variational approach to
the Sherrington-Kirkpatrick model. We propose here an iterative
procedure to show that, in the so-called Boltzmann ROSt,
Aizenman-Contucci (AC) polynomials naturally arise for almost
all values of the inverse temperature (not in average over
some interval only). The same results can be obtained
in any Quasi-Stationary ROSt, including therefore the Parisi structure.
The AC polynomials
impose restrictions on the overlap fluctuations in agreement with
Parisi theory.
\end{abstract}

\section{Introduction}

The study of mean field spin glasses has been very challenging from
both a physical and a mathematical point of view. It took several
years after the main model (the Sherrington-Kirkpatrick, or simply
SK) was introduced before Giorgio Parisi was able to compute the
free energy so ingeniously (\cite{mpv} and references therein), and
it took much longer still until a fully rigorous proof of Parisi's
formula was found \cite{guerra1, t4}. Parisi went beyond the
solution for the free energy and gave an Ansatz about the pure
states of the model as well, prescribing the so-called ultrametric
or hierarchical organization of the phases (\cite{mpv} and
references therein). From a rigorous point of view, the closest the
community could get so far to ultrametricity are identities
constraining the probability distribution of the {\em overlaps},
namely the Aizenman-Contucci (AC) and the Ghirlanda-Guerra
identities (see \cite{ac, gg} respectively). For further reading, we
refer to \cite{cg1, cg2, parisi5}, but also to the general
references \cite{talabook, bovierbook}. Most of the few important
rigorous results about mean field spin glasses can be elegantly
summarized within a powerful and physically profound approach
introduced recently by Aizenman et al. in \cite{ass}. We want to
show here that in this framework the AC identities can be deduced
too. This is achieved by studying a stochastic stability of some
kind, similarly to what is discussed in \cite{cg2}, inside the
environment (the {\em Random Overlap Structure}) suggested in
\cite{ass}, and taking into account also the intensive nature of the
internal energy density. A central point of the treatment is a power
series expansion similar to the one performed in \cite{barra}.

The paper is organized as follows.
In section 2 we introduce the concept of Random Overlap Structure
(henceforth ROSt), and use it to state
the related Extended Variational Principle.
In section 3 we present the main results regarding
the AC identities and similar families of relations.
In Appendix A we emphasize that the same results are valid in 
any Quasi-Stationary ROSt,
not just the Boltzmann one.


\section{Model, notations, previous basic results}

The Hamiltonian of the SK model
is defined on Ising spin configurations $\sigma:i\to\sigma_i=\pm 1$
of $N$ spins, labeled by $i=1,\ldots, N$, as
\begin{equation*}
\label{SK}
H_N(\sigma;J)=-\frac1{\sqrt N} \sum_{i<j}^{1,N} J_{ij} \sigma_i\sigma_j
\end{equation*}
where ${J_{ij}}$ are i.i.d. centered unit Gaussian random variables.
We will assume there is no external field.
Being a centered Gaussian
variable, the Hamiltonian is
determined by its covariance
$$
\mathbb{E}[H_N(\sigma)H_N(\sigma^{\prime})] =\frac{1}{2}Nq^2_{\sigma\sigma'}
$$
where
$$
q_{\sigma\sigma^{\prime}}=\frac{1}{N}\sum_{i=1}^N\sigma_i\sigma^{\prime}_i
$$
is the overlap, and $\mathbb{E}$ denotes here the expectation with respect to all
the (quenched) Gaussian variables.

The partition function $Z_N(\beta)$, the quenched free
energy density $f_N(\beta)$ and pressure $\alpha_N(\beta)$
are defined as:
\begin{eqnarray*}
\label{z}
 Z_N(\beta)&=& \sum_{\sigma}\exp(-\beta H_N(\sigma))\ , \\
\label{f}
-\beta f_N(\beta)&=&\frac{1}{N}\mathbb{E}\ln Z_N(\beta)=\alpha_N(\beta)\ .
\end{eqnarray*}
The Boltzmann-Gibbs average of an observable $\mathcal{O}(\sigma)$ is
denoted by $\omega$ and defined as
\begin{equation*}
\omega(\mathcal{O})=
Z_N(\beta)^{-1}
\sum_{\sigma}\mathcal{O}(\sigma)\exp(-\beta
H_N(\sigma))\ ,
\end{equation*}
but we will use the same $\omega$ to indicate in general
(weighted) sums over spins or non-quenched variables,
to be specified when needed, and with $\Omega$ we will
mean the product ({\sl replica}) measure of the 
needed number of copies of $\omega$.

Let us now introduce an auxiliary system.
\begin{definition}
A \emph{Random Overlap Structure}
$\mathcal{R}$ is a triple
$(\Sigma, \tilde{q}, \xi)$ where
\begin{itemize}
  \item $\Sigma\ni\gamma$ is a discrete space (set of abstract spin-configurations);
  \item $\tilde{q}:\Sigma^2\rightarrow[0, 1]$
  is a positive definite kernel (\emph{Overlap Kernel}), with 
  $|\tilde{q}|\leq 1$ (and $\tilde{q}=1$ on the diagonal of $\Sigma^{2}$);
  \item $\xi: \Sigma\rightarrow\mathbb{R}_+$
  is a normalized discrete positive random measure, i.e. a 
  system of random weights such that there is a probability measure $\mu$ on
  $[0,1]^{\Sigma}$ so that
  $\sum_{\gamma\in\Sigma}\xi_\gamma<\infty$ almost surely in the $\mu$-sense.
\end{itemize}
\end{definition}
The randomness in the weights $\xi$ is independent
of the randomness of the quenched variables from the
original system with spins $\sigma$.
We equip a ROSt with
two families of independent and centered Gaussians
$\tilde{h}_i$ and $\hat{H}$
with covariances
\begin{eqnarray}
\label{tilde}
\mathbb{E}[\tilde{h}_{i}(\gamma)\tilde{h}_{j}(\gamma^{\prime})]&=&
\delta_{ij}\tilde{q}_{\gamma\gamma^{\prime}}\ , \\
\label{hat}
\mathbb{E}[\hat{H}(\gamma)\hat{H}(\gamma^{\prime})]&=&
\tilde{q}^2_{\gamma\gamma^{\prime}}\ .
\end{eqnarray}
Given a ROSt $\mathcal{R}$ we define the trial pressure as
\begin{equation}
\label{trial}
G_N(\mathcal{R})=\frac{1}{N}\mathbb{E}
\ln\frac{\sum_{\sigma,\gamma}\xi_{\gamma}\exp(-
\beta\sum_{i=1}^N\tilde{h}_i(\gamma)\sigma_i)}
{\sum_{\gamma}\xi_{\gamma}\exp(-\beta\sqrt{\frac {N}{2}}
\hat{H}(\gamma))}\ ,
\end{equation}
where $\mathbb{E}$ denotes hereafter the expectation with respect
to all the (quenched) random variables (including the randomness in 
the random weights $\xi$) but spins $\sigma$ and
the abstract spins $\gamma$, the sum over which is in fact written
explicitly.
 
The following theorem (\cite{ass}) can be easily proven by interpolation
\begin{theorem}[Extended Variational Principle]
Infimizing for each $N$
separately the trial function $G_N(\mathcal{R})$ defined
in (\ref{trial}) over the whole
ROSt space, the resulting sequence tends to the limiting pressure
$-\beta f(\beta)$ of the SK model as $N$ tends to infinity
\begin{equation*}\label{}
 \alpha(\beta)\equiv\lim_{N\to\infty}\alpha_N(\beta)=\lim_{N\to\infty}
\inf_{\mathcal{R}}G_N(\mathcal{R}) \ .
\end{equation*}
\end{theorem}
For a given ROSt, 
the trial pressures \{$G_{N}$\} are a well defined sequence 
of real numbers indexed by $N$;
a ROSt $\mathcal{R}$ is said to be optimal if $\alpha\equiv
\lim_{N\to\infty}\alpha_{N}(\beta)=
\lim_{N\to\infty}G_N(\mathcal{R})\ \forall \beta\ .$
See \cite{arguin} for comments on some topological aspects
of the ROSt theory.
The space of all ROSt can be restricted to those ROSt's
enjoying some factorization property 
that all optimal ROSt's enjoy, without
missing the exact pressure \cite{g, lds1}. We will
therefore limit ourselves to these ROSt's, called 
Quasi-Stationary.

An optimal ROSt is the Parisi one (\cite{mpv, t4}),
another optimal one is the so-called Boltzmann ROSt $\mathcal{R}_B$,
defined as follows.
Take $\Sigma=\{-1,1\}^M$, and denote by $\tau$ the points of $\Sigma$.
We clearly have in mind an auxiliary spin systems (and that is
why we use $\tau$ as opposed to the previous $\gamma$ to
denote its points). In fact,
we also choose
\begin{equation*}
\tilde{h}_i=-\frac{1}{\sqrt{M}}\sum_{k=1}^M \tilde{J}_{ik}\tau_k\ ,\ \ \
\hat{H}=-\frac1M \sum_{k,l}^{1,M}\hat{J}_{kl}\tau_k\tau_l
\end{equation*}
which satisfy (\ref{tilde})-(\ref{hat}) with
$\tilde{q}_{\tau\tau^{\prime}}=\frac1M\sum_k\tau_k\tau_k^{\prime}$,
and $\tilde{J}$ and $\hat{J}$ are families of i.i.d.
random variables independent of the original
couplings $J$, with whom they share the same distribution
(i.e. all the $\tilde{J}$ and $\hat{J}$ are centered unit Gaussian random variables).
The variables $\tilde{h}_{.}$ are called {\sl cavity fields}.
Let us also choose
$$
\xi_\tau=\exp(-\beta H_M(\tau;\hat{J}))=
\exp\bigg(\beta\frac{1}{\sqrt{M}}\sum_{k,l}^{1,M}\hat{J}_{kl}\tau_{k}\tau_{l}\bigg)\ .
$$

If we call $\mathcal{R}_B(M)$ the structure defined above, we will formally
write $\mathcal{R}_B(M)\to\mathcal{R}_B$ as $M\to\infty$, and
we call $\mathcal{R}_B$ the Boltzmann ROSt.
The reason why such a ROSt is optimal is purely thermodynamic,
and equivalent to the existence of the thermodynamic limit of the
free energy per spin. A detailed proof of this fact can be found
in \cite{ass}; here we just mention the main point:
\begin{equation*}
\label{ }
\alpha(\beta)=\mathbf{C}\lim_M\frac1N\mathbb{E}\ln\frac{Z_{N+M}}{Z_M}
=\lim_{N\rightarrow\infty}\mathbf{C}\lim_MG_{N}(\mathcal{R}_{B}(M))
=G_{N}(\mathcal{R}_{B})=G(\mathcal{R}_{B})
\end{equation*}
where $\mathbf{C}\lim$ is the limit in the Ces{\`a}ro sense.
Notice that the Boltzmann ROSt does not depend on $N$, after the
$M$-limit.

\section{Analysis of the Boltzmann ROSt}

In this section we show that in the optimal Boltzmann ROSt's the
overlap fluctuations obey some restrictions, namely those
found by Aizenman and Contucci in \cite{ac}.
In other words we exhibit a recipe to generate the AC polynomials
within the ROSt approach.

\subsection{the internal energy term}\label{internal}

Let us focus on the denominator of the trial
pressure $G(\mathcal{R}_B)$, defined in (\ref{trial}),
computed at the Boltzmann ROSt $\mathcal{R}_B$,
defined in the previous section. Let us normalize this quantity
by dividing by $Z_N$ and weight $\hat{H}$ with an independent variable
$\beta^{\prime}$ as opposed to $\beta$, which appears in the
{\sl Boltzmannfaktor} $\xi_{\tau}$. As in the Boltzmann structure
we have actual spins ($\tau$) and we do not
use the spins $\sigma$ here, we will still use $\omega$ (or $\Omega)$
to denote the Boltzmann-Gibbs (replica) 
measure (at inverse temperature $\beta$) in the space $\Sigma=\{-1,1\}^M$.
Moreover, we will use the notation $\langle\cdot\rangle=\mathbb{E}\Omega(\cdot)$ 
and, if present, a subscript $\beta$
recalls that the {\sl Boltzmannfaktor} in $\Omega$ has inverse temperature
$\beta$.
More precisely, we are computing the left hand side of the next
equality to get this
\begin{lemma}\label{lisboa}
\begin{equation}
\label{deng}
\frac{1}{N}\mathbb{E}\ln
\Omega\exp\left(-\beta^{\prime}\sqrt{\frac {N}{2}}
\hat{H}(\tau)\right)=\frac{\beta^{\prime 2}}{4}(1-\langle \tilde{q}^2\rangle_{\beta})\ .
\end{equation}
\end{lemma}
Similar calculations have
been performed already, but in this specific context
the result has been only stated without proof in \cite{g}, while
a detailed proof is given only in the dilute case in \cite{lds1}.
So let us prove the lemma.
Let us take $M$ finite. Thanks to the property of addition of independent
Gaussian variables, the left hand side of (\ref{deng}) is
the same as
$$
\frac 1N\mathbb{E}\ln\frac{Z_{M}(\beta^{*})}{Z_{M}(\beta)}=
\frac MN (\alpha_{M}(\beta^{*})-\alpha_{M}(\beta))\ ,\
\beta^{*}=\sqrt{\beta^{2}+\frac{\beta^{\prime 2}N}{M}}
$$
which in turn, thanks to the convexity of $\alpha$, can be estimated as follows
$$
\frac MN(\beta^{*}-\beta)\alpha_{M}^{\prime}(\beta)
\leq\frac MN (\alpha_{M}(\beta^{*})-\alpha_{M}(\beta))
\leq\frac MN(\beta^{*}-\beta)
\alpha_{M}^{\prime}(\beta^{*})\ .
$$
Now
$$
\frac MN(\beta^{*}-\beta)=\frac{\beta^{\prime 2}}{2 \beta}+o(\frac{1}{M})\ ,\
\alpha^{\prime}(\beta)=\frac{\beta}{2}(1-\langle \tilde{q}^{2}\rangle_{\beta})\ .
$$
Therefore, when $M\to\infty$, we get (\ref{deng}) for almost all $\beta$, i.e.
whenever $\alpha^{\prime}(\beta^{*})\to\alpha^{\prime}(\beta)$, or equivalently
whenever $\langle\cdot\rangle_{\beta^{*}}\to\langle\cdot\rangle_{\beta}$.
Notice that the quantity in (\ref{deng}) does not depend on $N$ \cite{g, lds1}.
\begin{theorem}\label{acinternal}
The following statements hold:
\begin{itemize}
\item The left hand side of (\ref{deng}) is intensive
(does not depend on $N$);
\item The left hand side of (\ref{deng}) is a
monomial of order two in $\beta^{\prime}$;
\item The Aizenman-Contucci identities hold.
\end{itemize}
\end{theorem}
{\bf Proof}\\
Recall that $\hat{H}$ is a centered Gaussian, and so is therefore $-\hat{H}$
and the Gibbs measure is such that the substitution
$\hat{H} \to \hat{H} - \hat{H}^{\prime}$ implies
\begin{equation*}
\frac{1}{N}\mathbb{E}\ln\Omega
\exp \left(-\beta^{\prime}\sqrt{\frac{N}{2}}\hat{H}\right)=
\frac{1}{2N}\mathbb{E}\ln\Omega\exp \left(-\beta^{\prime}\sqrt{\frac{N}{2}}
(\hat{H}-\hat{H}^{\prime})\right) \ .
\end{equation*}
Expand now in powers of $\beta^{\prime}$ the exponential first and then
the logarithm.
\begin{multline*}
\frac{1}{N}\mathbb{E}\ln\Omega
\exp\left(-\beta^{\prime}\sqrt{\frac{N}{2}}\hat{H}\right)=
\frac{\beta^{\prime 2}}{4}(1-\langle \tilde{q}^2\rangle)= \\
\frac{1}{2N}\mathbb{E}\ln\Omega\left[1+
\frac{\beta^{\prime 2}}{2}\frac{N}{2}(\hat{H}-\hat{H}^{\prime})^2+
\frac{\beta^{\prime 4}}{4!}\frac{N^2}{4}(\hat{H}-\hat{H}^{\prime})^4
+\cdots\right]= \\
\frac{1}{2N}\mathbb{E}\left\{\left[\frac{N
\beta^{\prime 2}}{4}(2\Omega(\hat{H}^2)-2\Omega^2(\hat{H}))\right]+\right.\\
\frac{N^2}{4}\frac{\beta^{\prime 4}}{4!}\left[2\Omega(\hat{H}^4)-
8\Omega(\hat{H})\Omega(\hat{H}^3)+6\Omega^2(\hat{H}^2)\right] -\\
\left.\frac{N^2}{2}\frac{\beta^{\prime 4}}{4}\left[\Omega^2(\hat{H}^2)
+\Omega^4(\hat{H})-2\Omega(\hat{H}^2)\Omega^2(\hat{H})\right]+\cdots\right\}\ .
\end{multline*}
A straightforward calculation yields
$$
\mathbb{E}\Omega(\hat{H}^4) = 3\ ,  \ \ \ \
\mathbb{E}[\Omega(\hat{H}^3)\Omega(\hat{H})]= 3\langle
\tilde{q}_{12}^2\rangle\ ,  \ \ \ \
\ \mathbb{E}\Omega^2(\hat{H}^2) =
1 + 2\langle  \tilde{q}_{12}^4\rangle\ ,
$$
$$
\mathbb{E}[\Omega(\hat{H}^2)\Omega^2(\hat{H})]
= \langle  \tilde{q}_{12}^2\rangle + 2\langle  \tilde{q}_{12}^2
\tilde{q}_{13}^2\rangle\ , \ \ \ \
\mathbb{E}\Omega^4(\hat{H})=3\langle  \tilde{q}_{12}^2
\tilde{q}_{34}^2\rangle\
$$
and so on. All quantities of this sort can be computed in
the same way.
As an example, let us calculate
$\mathbb{E}[\Omega(\hat{H}^2)\Omega^2(\hat{H})]
=\mathbb{E}[\omega(\hat{H}^{2}_1)
\omega(\hat{H}_2)\omega(\hat{H}_3)]$. Like for overlaps,
subscripts denote replicas.
In order to evaluate the expectation of products of
Gaussian variables, we can use Wick's theorem: we just count
all the possible ways to contract the four Gaussian terms
$\hat{H}_1,\hat{H}_1,\hat{H}_2,\hat{H}_3$ and sum over
every non-vanishing contribution
\begin{eqnarray}
\langle \contraction{}{A}{C}{} \hat{H}_1 \hat{H}_2
 \contraction{}{A}{C}{} \hat{H}_1 \hat{H}_3 \rangle
 &=&
 \langle \tilde{q}^2_{12}\tilde{q}^2_{23}\rangle\ , \label{contr1} \\
\langle \contraction{}{A}{C}{} \hat{H}_1 \hat{H}_1
 \contraction{}{A}{C}{} \hat{H}_2 \hat{H}_3 \rangle
 &=& \langle 1 \cdot \tilde{q}^2_{12}\rangle\ , \label{contr2} \\
\langle \contraction{}{A}{C}{} \hat{H}_1 \hat{H}_3
 \contraction{}{A}{C}{} \hat{H}_1 \hat{H}_2 \rangle
 &=& \langle \tilde{q}^2_{12}\tilde{q}^2_{23}\rangle\ . \label{contr3}
 \end{eqnarray}
The sum of all the terms gives the exactly
$\langle \tilde{q}^2_{12} \rangle
+2\langle \tilde{q}^2_{12}\tilde{q}^2_{23}\rangle$.
Now equation (\ref{deng}) is therefore expressed in terms of an identity
for all $\beta^{\prime}$
of two polynomials in $\beta^{\prime}$: one is of order two, the other is a
whole power series. We can then equate the coefficients of same order,
or equivalently put to zero all the terms of order higher that two in $\beta^{\prime}$.
The consequent equalities are exactly the Aizenman-Contucci ones (\cite{ac}),
an example of these is
\begin{equation*}
\langle  \tilde{q}_{12}^4\rangle -4\langle
\tilde{q}_{12}^2 \tilde{q}_{13}^2\rangle +3\langle
\tilde{q}_{12}^2 \tilde{q}_{34}^2\rangle = 0\ ,
\end{equation*}
which arises from the lowest order in the expansion above.  $\Box$

\subsection{the entropy term}

In the same spirit as in the previous section,
let us move on to the normalized numerator of the trial pressure
$G(\mathcal{R}_B)$, defined in (\ref{trial}),
computed at the Boltzmann ROSt $\mathcal{R}_B$,
defined in the previous section.  If we define
\begin{equation*}
\label{ }
c_{i}=2\cosh(-\beta\tilde{h}_{i})=
\sum_{\sigma_{i}}\exp(-\beta \tilde{h}_{i}\sigma_{i})\ ,
\end{equation*}
then
\begin{equation}
\label{numg}
\frac{1}{N}\mathbb{E}
\ln\Omega\sum_{\sigma}\exp(-
\beta\sum_{i=1}^N\tilde{h}_i\sigma_i)=
\frac{1}{N}\mathbb{E}
\ln\Omega(c_1\cdots c_{N})
\end{equation}
does not depend on $N$ \cite{g, lds1}, if we consider
the infinite Boltzmann ROSt, where $M\to\infty$.

Again, assume we replace the $\beta$ in front of the cavity
fields $\tilde{h}_{.}$ (but not in the state $\Omega$)
with a parameter $\sqrt{t}$, and define, upon rescaling,
\be\label{psi}
\Psi(t)=\mathbb{E}\ln\Omega\sum_{\sigma}\exp
\frac{\sqrt{t}}{\sqrt{N}}\sum_{i=1}^{N}\tilde{h}_i\sigma_i\ .
\ee
We want to study the flux (in $t$) of equation
(\ref{psi}) to obtain an integrable expansion.
The $t$-flux of the cavity function $\Psi$ is given by
\begin{equation}\label{stream}
\partial_t\Psi(t)= \frac{1}{2}(1-\langle  q_{12}\tilde{q}_{12}\rangle _t)\ ,
\end{equation}
which is easily seen by means of a standard use of
Gaussian integration by parts. The subscript in $\langle \cdot \rangle_{t}
=\mathbb{E}\Omega_{t}$
means that such an average includes the $t$-dependent exponential
appearing in (\ref{psi}), beyond the sum over $\sigma$.
\begin{theorem} Let $F_{s}$ be measurable with respect to
the $\sigma$-algebra
generated by the overlaps of $s$ replicas of $\{\sigma\}$ and $\{\tau\}$.
Then the cavity streaming equation is
\begin{multline}
\partial_t\langle  F_s\rangle _t=\label{simmetria} \\
\langle  F_{s}(\sum_{\gamma,\delta}^{1,s}
q_{\gamma,\delta}\tilde{q}_{\gamma,\delta}
-s\sum_{\gamma=1}^{s}q_{\gamma,s+1}\tilde{q}_{\gamma,s+1} +
\frac{s(s+1)}{2}q_{s+1,s+2}\tilde{q}_{s+1,s+2})\rangle _t\ .
\end{multline}
\end{theorem}
{\bf Proof}\\
We consider the Boltzmann ROSt $\mathcal{R}_{B}(M)$ with any value of $M$.
The proof relies on the repeated application
of the usual integration by parts formula for Gaussian variables
\begin{eqnarray*}
\partial_t\langle  F_s\rangle _t &=& \partial_t
\mathbb{E}\frac{\sum_{\sigma\tau}F_s
  \exp({-\beta H_{M}(\tau)}) \exp({\sqrt{\frac{t}{MN}} \sum_{ij} \sum_{\gamma}
  \tilde{J}_{ij}
      \tau_i^{\gamma} \sigma_j^{\gamma}})}{\sum_{\sigma\tau}
  \exp({-\beta H_{M}(\tau)}) \exp({\sqrt{\frac{t}{MN}}}  \sum_{ij} \sum_{\gamma}
  \tilde{J}_{ij}
      \tau_i^{\gamma} \sigma_j^{\gamma})}\\
&=&\frac{1}{2\sqrt{tMN}}\mathbb{E}\sum_{ij}J_{ij}
\sum_{\gamma}^s(\Omega_{t}[F_s\tau_i^{\gamma}\sigma_j^{\gamma}]
-\Omega_{t}[F_s]\Omega_{t}[\tau_i^{\gamma}\sigma_j^{\gamma}])\\
&=&\frac{1}{2\sqrt{tMN}}\sum_{ij}\mathbb{E}J_{ij}
(\sum_{\gamma}\Omega_{t}[F_s\tau_i^{\gamma}\sigma_j^{\gamma}]
- s\Omega_{t}[F_s]\omega[\tau_i\sigma_j])\\
&=&\frac{1}{2MN}\sum_{ij}\mathbb{E}(\sum_{\gamma,\delta}
\Omega_{t}[F_s\sigma_j^{\gamma}\tau_i^{\gamma}
\sigma_j^{\delta}\tau_i^{\delta}]\\
&{}&-\sum_{\gamma,\delta}\Omega_{t}
[F_s\tau_i^{\gamma}\sigma_j^{\gamma}]
\Omega_{t}[\tau_i^{\delta}\sigma_j^{\delta}]-
s\omega_t[\tau_i\sigma_j]\sum_{\delta}(\Omega_{t}
[F_s\tau_i^{\delta}\sigma_j^{\delta}]\\
&{}&-\Omega_{t}[F_s]\Omega_{t}
[\tau_i^{\delta}\sigma_j^{\delta}] -s
\Omega_{t}[F_s](1-\omega^2_t[\tau_i\sigma_j])))\\
&=&\frac{1}{2}\mathbb{E}
(\sum_{\gamma,\delta}\Omega_{t}
[F_sq_{\gamma,\delta}\tilde{q}_{\gamma,\delta}] - s
\sum_{\gamma}\Omega_{t}
[F_sq_{\gamma,s+1}\tilde{q}_{\gamma,s+1}]\\
&{}& +ss\Omega_{t}[F_s q_{s+1,s+2}\tilde{q}_{s+1,s+2} ] \\
&{}&-s\Omega_{t}[F_s]\Omega_{t}
[F_s  q_{s+1,s+2}\tilde{q}_{s+1,s+2} ])\ ,
\end{eqnarray*}
where in $\Omega_{t}$ we have included the sum over $\sigma$ and $\tau$,
the {\sl Boltzmannfaktor} in $\tau$, and the $t$-dependent exponential.
At this point, remembering that $\tilde{q}_{\gamma\gamma}=1$, we can write
$$
\sum_{\gamma,\delta}\Omega_{t}
[F_s q_{\gamma\delta} \tilde{q}_{\gamma\delta} ] = 2
\sum_{\gamma,\delta}\Omega_{t}
[F_s  q_{\gamma\delta} \tilde{q}_{\gamma\delta}] + s \Omega_{t}[F_s]\ .
$$
which completes the proof. $\Box$

Now the way to proceed is simple: we have to expand the
$t$-derivative of $\Psi(t)$ (right hand side of (\ref{stream})) using the cavity
streaming equation (\ref{simmetria}), and
we will stop the iteration at the first non trivial order 
(that is expected to be at least four, being the 
first AC relation of that order). Once a
closed-form expression is in our hands, we can write down
an order by order expansion of
the (modified) denominator of the Boltzmann ROSt
(that is the function $N^{-1}\psi(t)$ evaluated for $t=N\beta^2$).
We have
\begin{equation*}\label{capo}
\partial_t\langle  q_{12}\tilde{q}_{12}\rangle _t =
\langle    q_{12}^2\tilde{q}_{12}^2 -4 q_{12}\tilde{q}_{12}q_{23}\tilde{q}_{23}
+3 q_{12}\tilde{q}_{12}q_{34}\tilde{q}_{34}\rangle _t\ .
\end{equation*}
After the first iteration:
\begin{equation*}
\partial_t\langle  q_{12}^2\tilde{q}_{12}^2\rangle _t =
\langle   q_{12}^3\tilde{q}_{12}^3 -
4 q_{12}^2\tilde{q}_{12}^2\tilde{q}_{23}p_{23}
+3 q_{12}^2\tilde{q}_{12}^2\tilde{q}_{34}\rangle _t \ ,
\end{equation*}
\begin{multline*}
 \partial_t\langle   \tilde{q}_{12}q_{12}\tilde{q}_{23}q_{23}\rangle _t =
\langle    \tilde{q}_{12}q_{12}\tilde{q}_{23}q_{23}\tilde{q}_{13}q_{13}+
2\tilde{q}_{12}^2q_{12}^2\tilde{q}_{23}q_{23}\\
-6 \tilde{q}_{12}q_{12}\tilde{q}_{23}q_{23}\tilde{q}_{34}q_{34} -
3 \tilde{q}_{12}q_{12}\tilde{q}_{13}q_{13}\tilde{q}_{14}q_{14}
 +6 \tilde{q}_{12}q_{12}\tilde{q}_{34}q_{34}\tilde{q}_{45}q_{45}\rangle _t \ ,
 \end{multline*}
 \begin{multline*}
 \partial_t\langle   \tilde{q}_{12}q_{12}\tilde{q}_{34}q_{34}\rangle _t =
\langle  4\tilde{q}_{12}q_{12}\tilde{q}_{23}q_{23}\tilde{q}_{34}q_{34}
+2\tilde{q}_{12}^2q_{12}^2\tilde{q}_{34}q_{34}\\
-16 \tilde{q}_{12}q_{12}\tilde{q}_{34}q_{34}\tilde{q}_{45}q_{45} +
10 \tilde{q}_{12}q_{12}\tilde{q}_{34}q_{34}\tilde{q}_{56}q_{56}\rangle _t\ .
\end{multline*}
The higher orders can be obtained exactly in the same way,
so we can write down right away the expression for
$\langle q_{12}\tilde{q}_{12}\rangle$,
referring to \cite{ac, barra} for a detailed
explanation of this iterative method:
\begin{multline}\label{expansion}
\langle  q_{12}\tilde{q}_{12}\rangle _t=
\langle  q_{12}^2\tilde{q}_{12}^2\rangle  t
-2 \langle  q_{12}\tilde{q}_{12}q_{23}\tilde{q}_{23}
q_{13}\tilde{q}_{13}\rangle  t^2-\frac{1}{6}\langle
q^4_{12}\tilde{q}^4_{12}\rangle  t^3\\
-2 \langle  q_{12}^2\tilde{q}_{12}^2q_{23}^2\tilde{q}_{23}^2\rangle  t^3
+\frac{3}{2}\langle  q_{12}^2\tilde{q}_{12}^2q_{34}^2\tilde{q}_{34}^2\rangle
t^3+6 \langle  q_{12}\tilde{q}_{12}q_{23}\tilde{q}_{23}
q_{34}\tilde{q}_{34}q_{14}\tilde{q}_{14}\rangle  t^3\ .
\end{multline}
Notice that the averages no longer depend on $t$.
In this expansion we considered both
$q$-overlaps and $\tilde{q}$-overlaps, but as
the sum over the spins $\sigma$ can be performed explicitly,
we can obtain an explicit expression at least for
the $q$-overlaps, and get
\begin{eqnarray*}
\langle  q^2_{12}\rangle
&=& \frac{1}{N^2}\mathbb{E}\sum_{ij}
\omega^2(\sigma_i\sigma_j)=\frac{1}{N}\ , \\
\langle q_{12}q_{23}q_{31}\rangle
&=& \frac{1}{N^3}\mathbb{E}\sum_{ijk}
\omega(\sigma_i\sigma_j)\omega(\sigma_j\sigma_k)
\omega(\sigma_k\sigma_i) = \frac{1}{N^2}\ , \\
\langle  q^2_{12}q^2_{34}\rangle
&=& \frac{1}{N^4}\mathbb{E}\sum_{ijkl} \omega^2(\sigma_i\sigma_j)
\omega^2(\sigma_k\sigma_l)=\frac{1}{N^2}\ , \\
\langle  q_{12}q_{23}q_{34}q_{14}\rangle
&=& \frac{1}{N^4}\mathbb{E}\sum_{ijkl}
\omega(\sigma_i\sigma_j)\omega(\sigma_j\sigma_k)
\omega(\sigma_k\sigma_l)\omega(\sigma_l\sigma_i) =\frac{1}{N^3}\ ,\\
\langle  q^4_{12}\rangle
&=& \frac{1}{N^4}\mathbb{E}\sum_{ijkl}
\omega(\sigma_i\sigma_j\sigma_k\sigma_l)
\omega(\sigma_i\sigma_j\sigma_k\sigma_l) = \frac{3(N-1)}{N^3} +
\frac{1}{N^3}\ ,\\
\langle  q^2_{12}q^2_{23}\rangle
&=&
\frac{1}{N^4}\mathbb{E}\sum_{ijkl}
\omega(\sigma_i\sigma_j)\omega(\sigma_i\sigma_j\sigma_k\sigma_l)
\omega(\sigma_i\sigma_j)=\frac{1}{N^2}\ .
\end{eqnarray*}
Moreover, as the $q$-overlaps have been
calculated explicitly, we can use a graphical formalism \cite{ac, barra}.
In such a formalism we use points to identify replicas and lines for the
overlaps between them. So for example:
$$
\langle  
\includegraphics{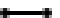} 
\rangle =\langle  \tilde{q}_{12}\rangle ,\qquad
\langle  \includegraphics{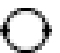}\rangle =\langle
\tilde{q}^2_{12}\rangle , \qquad \langle  \includegraphics{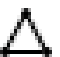}\rangle
=\langle  \tilde{q}_{12}\tilde{q}_{23}\tilde{q}_{13}\rangle
$$
and so on. Now we can integrate (\ref{stream}) thanks to the
polynomial expansion based on (\ref{expansion}) and to the
expressions for the $q$-fluctuations. We obtain
$$
\Psi(t)=\frac{1}{2}\int_0^t[1-\langle q_{12}\tilde{q}_{12}\rangle
_{t^{\prime}}]dt^{\prime}\ ,
$$
\begin{multline}\label{main}
\frac{1}{N}\Psi(t=N\beta^{2})=
\frac{\beta^{2}}{2}-\langle  \includegraphics{02a.eps}\rangle 
\frac{\beta^{4}}{4} +
\langle  \includegraphics{03b.eps}\rangle \frac{\beta^{6}}{3}-
\langle  \includegraphics{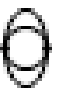}\rangle \frac{\beta^{8}}{24}-\\
\langle  \includegraphics{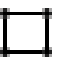}\rangle \frac{3\beta^{8}}{4}
+N\beta^{8}[ \frac{1}{16}\langle
\includegraphics{04g.eps}\rangle  - \frac{1}{4}\langle
\includegraphics{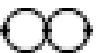}\rangle +\frac{3}{16}\langle
\includegraphics{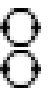}\rangle ]\ .
\end{multline}
This expression, though truncated at
this low order, already looks pretty much alike the expansion found
using the internal energy part of the Boltzmann pressure.

We stress however two important features of expression (\ref{main}).
The first is that within this approach we do not have problems 
concerning the Replica Simmetry Anzatz (RS) \cite{mpv}, 
and this can be seen by the proliferating of the overalaps 
fluctuations, via which we expand the entropy (a RS 
theory does not allow
such fluctuations). Secondly, we note that not all the terms 
inside the equations
(\ref{main}) are intensive: the last three graphs are all 
multiplied by a factor $N$.
Recalling that this expansion does not depend on $N$, 
and physically a density is intensive by definition, 
we put to zero all the terms in the squared bracket, so to have
\begin{equation*}
\langle  \includegraphics{04g.eps}-4\includegraphics{04h.eps}+3
\includegraphics{04f.eps}\rangle =0\ .
\end{equation*}
Again we can find the AC identities.


\appendix

\section{Extension to all Quasi-Stationary ROSt's}

For sake of simplicity, all the explicit calculation 
we performed took into account the 
Boltmann structure only. But the whole content actually does
not depend on the explicit form of the Hamiltonians, it merely relies
on the Gaussian nature of the random variables and their
moments, independently of the 
space they are defined in. In other words, as long as we consider 
centered Gaussian variables, the whole treatment depends only on their
covariances. That is why changing the ROSt does not change the results,
except the overlaps in the various expressions will be those of the considered
ROSt (e.g. the ultrametric Parisi trial overlaps), provided
some properties are preserved (Quasi-Stationarity).

Let us focus for instance on the internal energy part, which is simpler,
and see that the results of subsection \ref{internal} are the same in any 
Quasi-Stationary ROSt. 
First of all, notice that the proof of Theorem \ref{acinternal} never makes
use of the explicit form of the Hamiltonians and therefore 
(\ref{contr1})-(\ref{contr2})-(\ref{contr3}) stay identical, as they are
determined purely by the covariances of the Hamiltonians. The same
clearly holds for all the other terms not explicitly considered in the example.

So the validity of the results coincides with 
the validity of Lemma \ref{lisboa}.
The ROSt's for which such 
a lemma holds are called Quasi-Stationary (see \cite{ass2,arguin}),
in this case with respect to the Cavity Step 
(see \cite{g, ass2}).
Notice that the left hand side of (\ref{deng}) is zero for
$\beta^{\prime}=0$ independently of the particular ROSt. Hence by
the fundamental theorem of calculus the same left hand side
coincides with the integral from zero to $\beta^{\prime}$
of its derivative (with respect to $\beta^{\prime}$). But the
form of such a derivative is just determined by the covariance
of $\hat{H}$ (this is at the heart of 
\cite{ass}), which is always defined to be an overlap.
Therefore a simple Gaussian integration by parts, as illustrated
in \cite{ass}, leads to the right hand side of (\ref{deng}). These
are the intuitive reasons that heuristically explain why 
both the explicit calculation in Lemma \ref{lisboa} and the 
expansion of Theorem \ref{acinternal} are the same in any 
Quasi-Stationary ROSt,
and so are the AC polynomials, no matter what the overlap 
looks like in a generic abstract space.
So AC polynomials hold in any Quasi-Stationary ROSt, non-optimal too, 
but if the chosen ROSt is not optimal there will be no overlap locking and the
trial overlap will have very little to share with the true ones of the model.
Moreover (\ref{deng}) will not in general provide the internal energy of the model
(but this can be the case in some optimal ROSt too, like the Parisi one!).

\section*{Conclusions and Outlook}

We have shown how some constraints on the distribution of
the overlap naturally arise within the Random Overlap Structure
approach. As our analysis of the Boltzmann ROSt is similar
to the study of stochastic stability it is not surprising that the
constraints coincide with the Aizenman-Contucci identities.
In the ROSt context, such identities are easily connected
with the existence of the thermodynamic limit of the free
energy density (which is equivalent to the optimality of
the Boltzmann ROSt) and with the physical fact that
the internal energy is intensive. We also showed that,
as expected, the entropy part of the free energy yields
the same constraints as the other part (i.e. the internal energy).

The hope for the near future is that the ROSt approach will lead
eventually to a good understanding of the pure states and the
phase transitions of the model.
A first step has been taken in \cite{g}, and our
present results can be considered as
a second step in this direction. (Other more interesting results
regarding the phase transition at $\beta=1$ can also be obtained with
the same techniques employed here, including the graphical
representation \cite{abds}.) A further step should
bring the Ghirlanda-Guerra identities, and then hopefully a proof of
ultrametricity.

\section*{Acknowledgment}

The authors warmly thank Francesco Guerra and Pierluigi Contucci for
a precious scientific and personal support, and sincerely thank
Peter Sollich for useful discussions and much advice. A grateful appreciation
is owed to Louis-Pierre Arguin for pointing out a weak point 
in a previous version of the paper and for very fruitful conversations.

\end{document}